\documentclass[
    ,final            % use final for the camera ready runs
  ]
  {aipproc}

\layoutstyle{6x9}

\begin{document}

\title{Study of the ATLAS central calorimeters response to pions in 2004 combined test beam}

\classification{01.30.Cc; 29.40.Vj}
\keywords      {ATLAS; Testbeam; Calorimetry; Pions}

\author{V.~Giangiobbe}{
  address={Laboratoire de Physique Corpusculaire de Clermont-Ferrand}
}

\begin{abstract}
During summer 2004, a testbeam for the ATLAS experiment was performed at CERN.
A full slice of the ATLAS detector was tested with the final version of the
acquisition electronics. 
In the talk presented at Calor06
conference, the ATLAS calorimeters response to high and low energy pions 
was presented. 
\end{abstract}

\maketitle

%%%%%%%%%%%%%%%%%%%%%%%%%%%%%%%%%%%%%%%%%%%%
%% MAINMATTER
%%%%%%%%%%%%%%%%%%%%%%%%%%%%%%%%%%%%%%%%%%%%

\section{Description of the 2004 combined test beam of the ATLAS detector}

\subsection{Central calorimetry in ATLAS}
\begin{figure}[h]
  \includegraphics[height=.2\textheight]{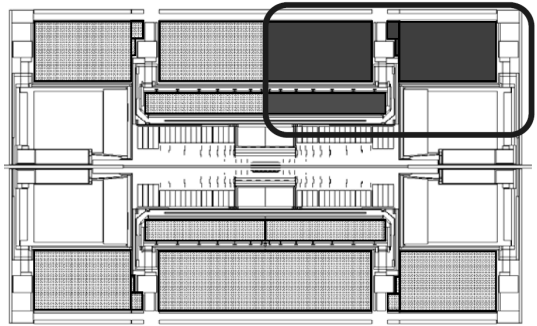}
  \caption{The calorimetry in ATLAS experiment.}
  \label{calo2d}
\end{figure}

The whole calorimetry in ATLAS is contained within a 12~m long cylinder with
an outer radius of about 4.2~m. This volume is divided into 3 part along
$\eta$. The central part called barrel, covers $-1.7<\eta<1.7$. The 
electromagnetic calorimeter (LAr calorimeter) uses lead as absorber and
the active material is liquid argon. For the hadronic calorimeter
(TileCal) the absorber is iron and the active part consists of
scintillating tiles. For mechanical reason the central hadronic calorimetry
is separated in 3 parts : 1 long barrel for $-1.0<\eta<1.0$ and 2 extended
barrels for $0.8<|\eta|<1.7$. Figure~\ref{calo2d} shows the
calorimetry in ATLAS. The central calorimetry which will be the subject of the
present document has a light-grey color. The dark-grey circled
region represents the part that have been exposed to the beam during the
2004 combined test-beam.   

   \subsection{Central calorimetry in 2004 combined test beam}
\begin{figure}[h]
  \includegraphics[height=.2\textheight]{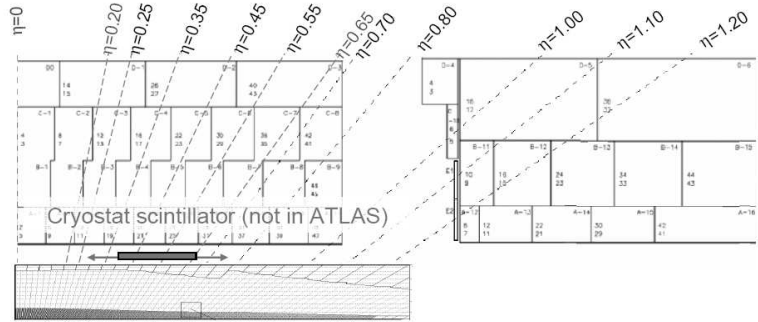}
  \caption{Top view of the calorimetry at the 2004 combined test-beam.}
  \label{CTB-top-view}
\end{figure}

\begin{figure}[h]
  \includegraphics[height=.2\textheight]{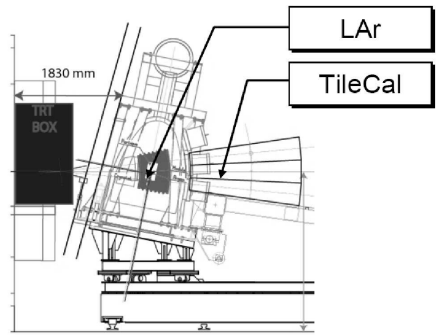}
  \caption{Side view of the calorimetry at the 2004 combined test-beam.}
  \label{CTB-side-view}
\end{figure}

The 2004 combined test-beam is the first test of a full ATLAS slice (inner
detector, calorimetry and muon spectrometer). In the following only the
calorimeter response will be discussed.
A top view of the calorimetry setup is shown
on Figure~\ref{CTB-top-view}. The LAr calorimeter is on the bottom.
Half a barrel of TileCal and 1 TileCal extended barrel are also shown. 
The modules were laying on a rotative Table giving the possibility to shoot
the beam at various $\eta$ positions (represented in
Figure~\ref{CTB-top-view}). Comparing with Figure~\ref{calo2d} it can be
observed that the test-beam geometry is similar to the
ATLAS one. Figure~\ref{CTB-side-view} shows a side view of the calorimetry
on the Table. We can see the LAr calorimeter within its large cryostat, and
3 TileCal modules. The $\phi$-coverage is 0.4 in LAr and 0.3 in TileCal.

   \subsection{Data sample and energy reconstruction}
During summer 2004, central ATLAS calorimeter modules have been exposed to muons, pions
and electron beams with different energies.
In the following only the results obtained using the pion beams are reported.
The High Energy (HE) pions had the following energies : 50, 100, 150, 180, 200,
250, 320 and 350~GeV. They had an impact angle with respect to the surface of
the detector corresponding to the pseudo-rapidity values : $\eta$=0.20, 0.25, 0.35,
0.45, 0.55, 0.65. The Very Low Energy (VLE) pion beams had energies ranging
between 1 and 9~GeV and the same $\eta$ values.

In LAr calorimeter the energy scale is obtained using the features of a
discharging capacitor. 
For TileCal, the energy scale is determined using the response of the
detector to electron beams. 
In the present analysis the energy that is deposited by pions in
calorimeters is reconstructed summing the energy of cells in a limited window around
the direction of the beam. It has been checked that the pion showers are
well contained in $\Delta\eta\times\Delta\phi=0.4\times0.4$ volume in LAr
and $\Delta\eta\times\Delta\phi=0.4\times0.3$ volume in LAr.   

\section{Pions that do not interact in the electromagnetic calorimeter}
Around 20\% of the high energy pions reach the hadronic calorimeter without
producing a shower in the electromagnetic section. These pions constitute
an interesting probe to check the performances of TileCal standalone and to
establish the absolute energy scale in TileCal. In
this section the response (ratio between the reconstructed and the beam
energy) and the resolution of TileCal will be reported for high energy
pions. 

   \subsection{Event selection}
\begin{figure}[h]
  \includegraphics[height=.2\textheight]{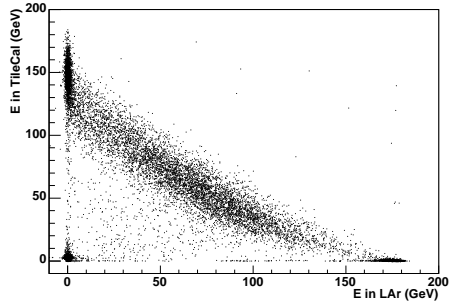}
  \caption{Correlation between LAr and TileCal energy}
  \label{ScatterPlot}
\end{figure}

\begin{figure}[h]
  \includegraphics[height=.2\textheight]{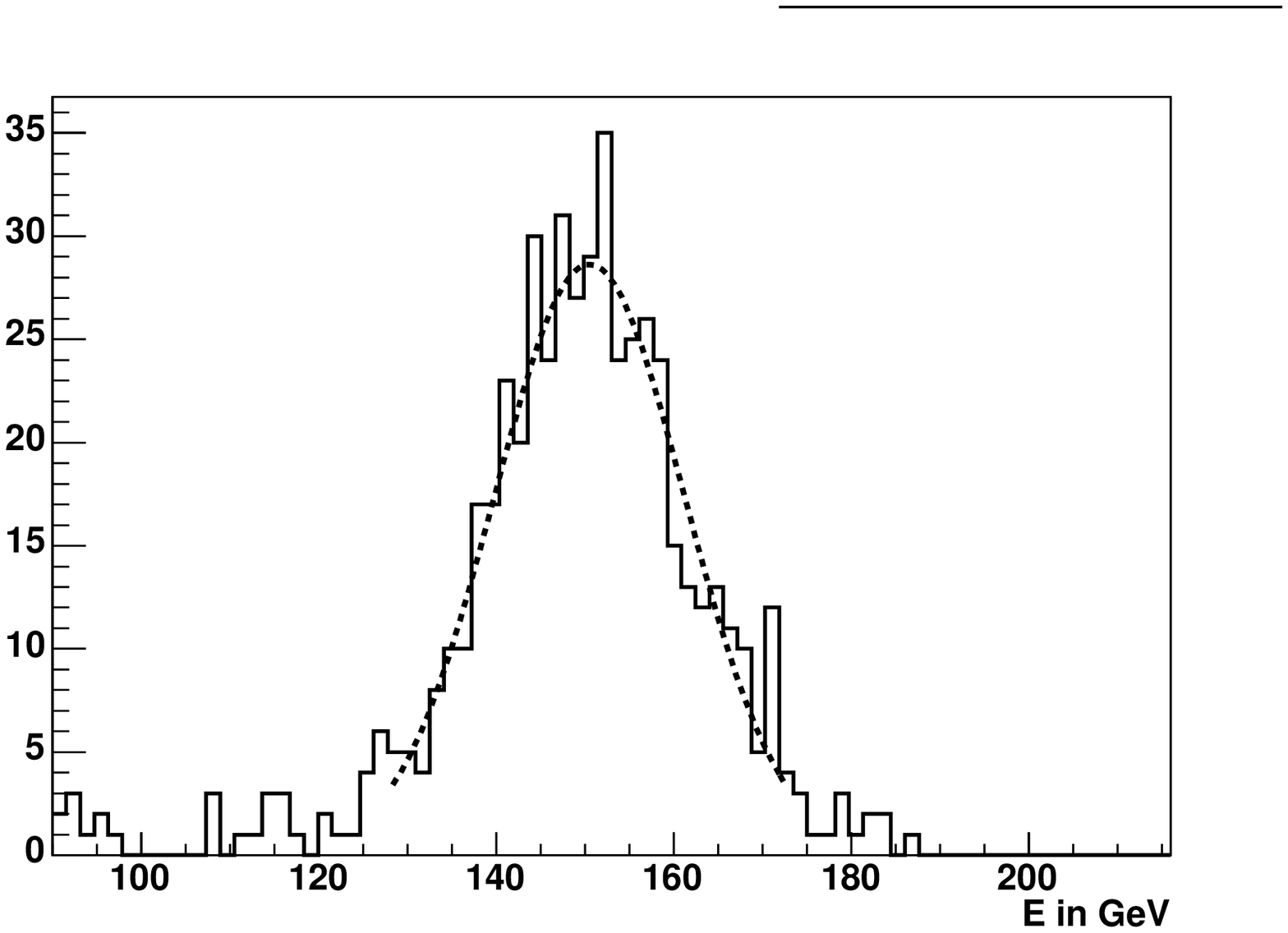}
  \caption{Response to 180~GeV pions that do not start showering
  before reaching TileCal.}
  \label{PionsSL}
\end{figure}

Figure~\ref{ScatterPlot} shows the correlation between the deposited energy
in the electromagnetic and hadronic calorimeters for a 180~GeV beam. It can
be seen on this plot that the high energy pion beam contains a significant
fraction of muons and electrons. Muons give a very small energy in both
calorimeters while electrons deposit all their energy in LAr. Muons and
electrons can be easily rejected looking at the longitudinal profile of the
energy in calorimeters. Pions showering in both calorimeters are visible in
the center of Figure~\ref{ScatterPlot} (anti-correlation between LAr and
TileCal). Pions interacting only in TileCal can be selected requesting a
signal compatible with the one of a minimum ionizing particle (mip) in each
of the 4 longitudinal compartments of the LAr. This is not enough to select
pions reaching TileCal without early interaction. Indeed pions passing
through the LAr at their minimum of ionization can interact in the outer
cryostat wall. In order to reject such events an additional scintillator
(called cryostat scintillator) has been placed between the outer cryostat
wall and the front face of TileCal (see Figure~\ref{CTB-top-view}). A
mip-like signal is then requested also in the cryostat scintillator.
Figure~\ref{PionsSL} shows the energy distribution in TileCal for 180~GeV
pions that pass the cuts described above (ie mip-like signal in each LAr
layer and in the cryostat scintillator). A Gaussian fit has been applied in
order to estimate the mean reconstructed energy $E_{raw}$ and the width
$\sigma$ of the distribution. The fit was performed in a region of 2 sigmas
around the peak value.  

%\begin{figure}[h]
%   \begin{center}
%      \begin{minipage}[c]{1.00\linewidth}         
%	 \begin{minipage}[t]{0.50\linewidth}
%            \begin{center}
%               \includegraphics[width=\linewidth]{Figures/ScatterPlot.eps}
%               \caption{Correlation between LAr and TileCal energy.}
%               \label{ScatterPlot}	    
%	    \end{center}
%         \end{minipage}
%	 
%	 \hspace{0.2in}
%	 
%	 \begin{minipage}[t]{0.50\linewidth}
%            \begin{center}
%               \includegraphics[width=\linewidth]{Figures/PionsSL.eps}
%               \caption{Response to 180~GeV pions that do not start showering
%	       before reaching TileCal.}
%	       \label{PionsSL}	    
%	    \end{center}
%         \end{minipage}      
%      \end{minipage}
%   \end{center}
%\end{figure} 

   \subsection{Response and resolution}
\begin{figure}[h]
  \includegraphics[height=.4\textheight]{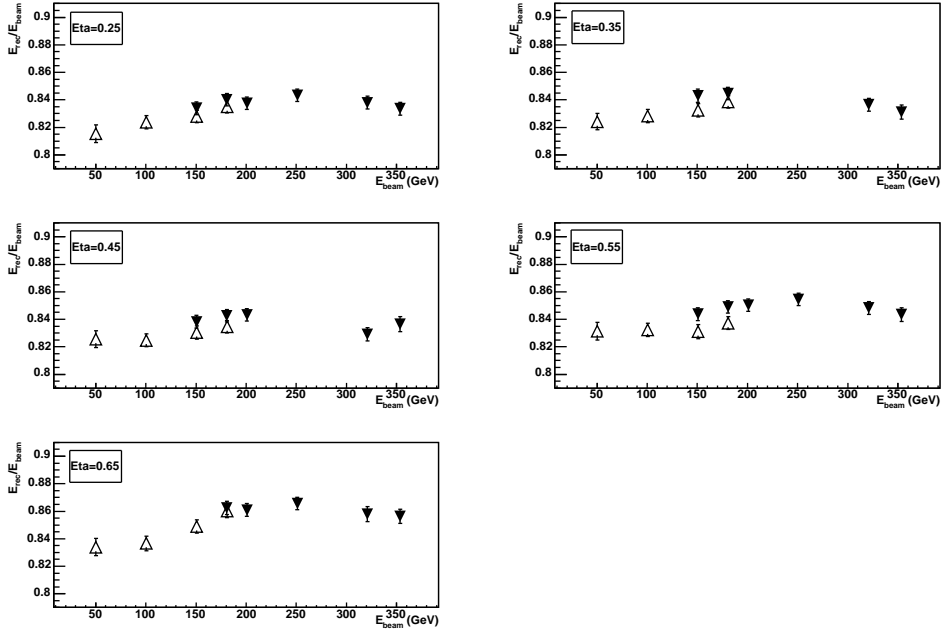}
  \caption{Ratio between the reconstructed energy and the beam energy for
  various $\eta$ positions. Black triangles represent negative beam results and
  white triangles represent positive beam ones (see text).}
  \label{Response-SL}
\end{figure}

Figure~\ref{Response-SL} shows the ratio between $E_{raw}$ and the 
beam energy for different $\eta$ positions. Two different beams have been
used during the data taking. One has a negative polarity and is expected to
contain only negative pions while the other one is a positive beam for
which a proton contamination is expected. It is a well known fact that
protons produce a lower response than pions in a non-compensating calorimeter
like TileCal \cite{TileCal-2001-005}. For this reason and because it was
not possible to separate pions and protons in 2004 combined test-beam, the
response for positive and negative beams have been represented separately
in Figure~\ref{Response-SL}. It can be noticed that, as expected, the
response for the positive beam, (150 and 180~GeV) is systematically
lower than the negative beam one at the same energy. The effect
is around 1\%. Looking at
Figure~\ref{Response-SL} it can be noticed that the calorimeter response is increasing
with the beam energy. This behavior is typical of a
non-compensating calorimeter. At high energy (320 and 350~GeV), due to the
transversal and longitudinal leakage, the response
decreases slightly. So far two sources of systematic effect have been
considered. The beam incident energy is known with a 0.5\% precision.
The influence of the selection cuts has been estimated to be of the
order of 0.2\%. This gives a global systematic uncertainty of around 0.5\%.
The statistical error on the reconstructed energy is around 1\%. It means
that the statistical error dominates the estimation of the TileCal
response to pions.

\begin{figure}[h]
  \includegraphics[height=.4\textheight]{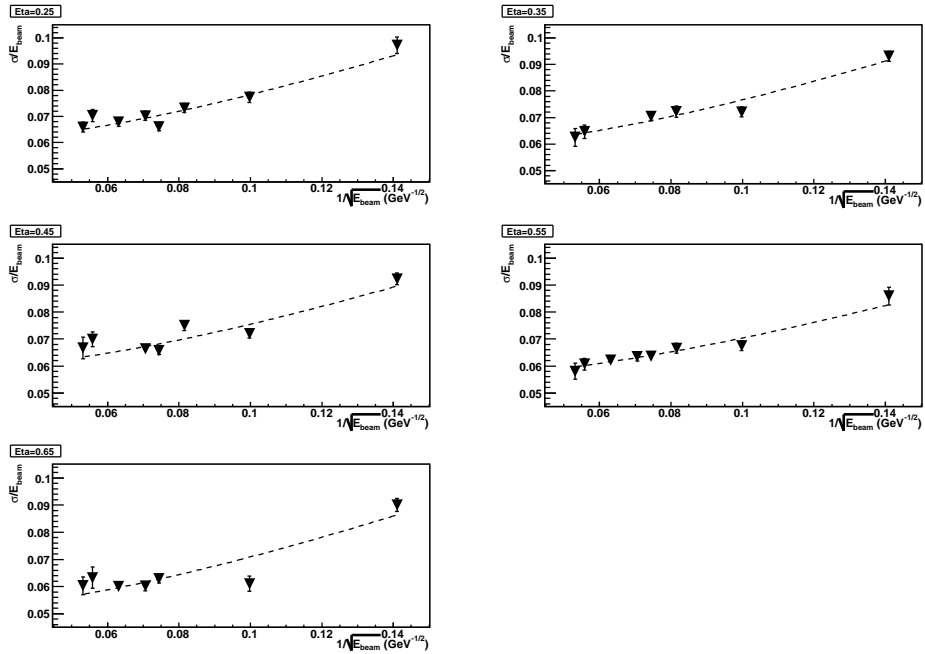}
  \caption{Resolution ($\sigma/E_{raw}$) as a function of $1/\sqrt{E_{beam} }$.}
  \label{Resolution-SL}
\end{figure}

The evolution of the resolution
defined as $\sigma/E_{raw}$ as a function of $1/\sqrt{E_{beam}}$ it shown in
Figure~\ref{Resolution-SL}. This study has been done using the negative 
beam (no protons) except at 50 and 100~GeV. A fit function has been
applied in order to estimate the sampling term $a$ and the constant term
$b$ :
$$ \frac{\sigma}{E_{raw}}=\frac{a}{\sqrt{E_{beam}}}\oplus b$$
The obtained parameters are given in Table~\ref{tab1}. It has to be
stressed that these results have been obtained without any correction for
leakage or non-compensation. For this reason the resolution is not the
best that can be achieved. The results of Table~\ref{tab1} are in
good agreement with results obtained in the same conditions during previous
TileCal standalone test-beams \cite{TILECAL-99-020}.
\begin{table}
\begin{tabular}{lrrrrr}
\hline
  & \tablehead{1}{r}{b}{Sampling term $a$\\($GeV^{1/2}$)}
  & \tablehead{1}{r}{b}{Constant term $b$\\(\%)} \\
\hline
$\eta=0.25$ & 0.51$\pm$0.03 & 5.9$\pm$0.2 \\
$\eta=0.35$ & 0.52$\pm$0.02 & 5.7$\pm$0.3 \\
$\eta=0.45$ & 0.48$\pm$0.02 & 5.8$\pm$0.2 \\
$\eta=0.55$ & 0.44$\pm$0.03 & 5.5$\pm$0.2 \\
$\eta=0.65$ & 0.49$\pm$0.03 & 5.1$\pm$0.1 \\
\hline
\end{tabular}
\caption{Sampling term $a$ and constant term $b$ for pions in TileCal.}
\label{tab1}
\end{table} 

\section{Combined calorimetry response to pions}
   \subsection{Energy reconstruction}
\begin{figure}[h]
  \includegraphics[height=.4\textheight]{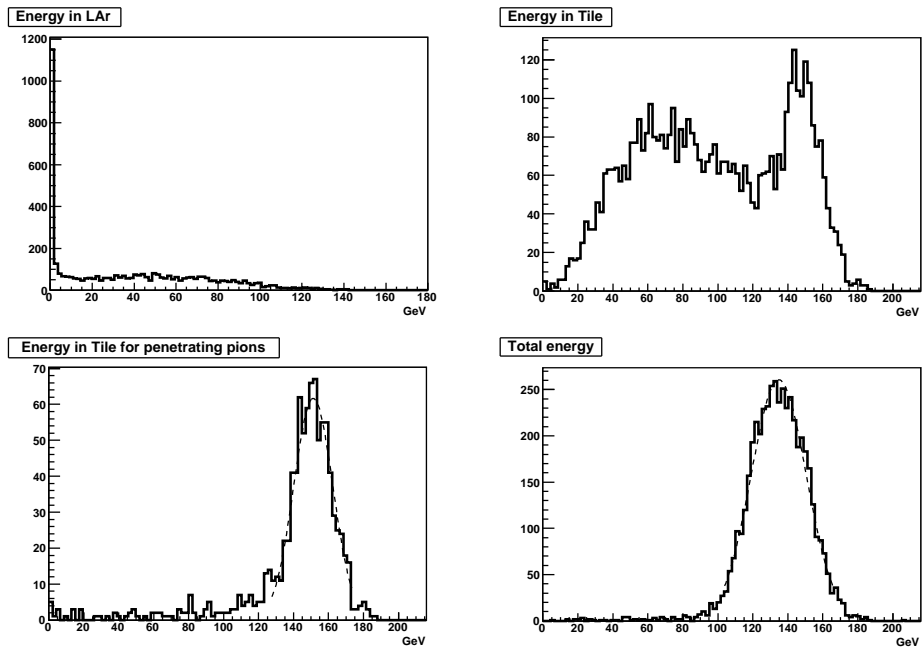}
  \caption{Energy distributions for 180~GeV pions at $\eta=0.35$
  Top~left~:~energy in LAr; top~right~:~energy in TileCal;
  bottom~left~:~energy in TileCal for pions that start showering in TileCal;
  bottom~right~:~energy in LAr+TileCal for all pions.}
  \label{Pions-combined}
\end{figure}

The study of combined calorimetry response to pions is done only using beams
with a negative polarity in order to get rid of any proton contamination. 
The energy in calorimeters is computed as the sum of energy
in LAr and energy in TileCal. No correction for non-compensation, dead material or
leakage has been applied. The aim of this analysis is to provide the 
combined response that can be used to fix the absolute hadronic scale and that
can be reproduced by the simulation.
Figure~\ref{Pions-combined} shows the energy distribution in LAr, TileCal,
and the sum of the energies in the two calorimeters for 180~GeV pions
at $\eta=0.35$. The reconstructed energy for pions that start showering in
TileCal (called penetrating pions) is also shown (bottom left). As before a
Gaussian fit is applied on the total energy distribution and the mean
reconstructed energy $E_{rec}$ is obtained from the fit. 

   \subsection{Combined response}
\begin{figure}[h]
  \includegraphics[height=.4\textheight]{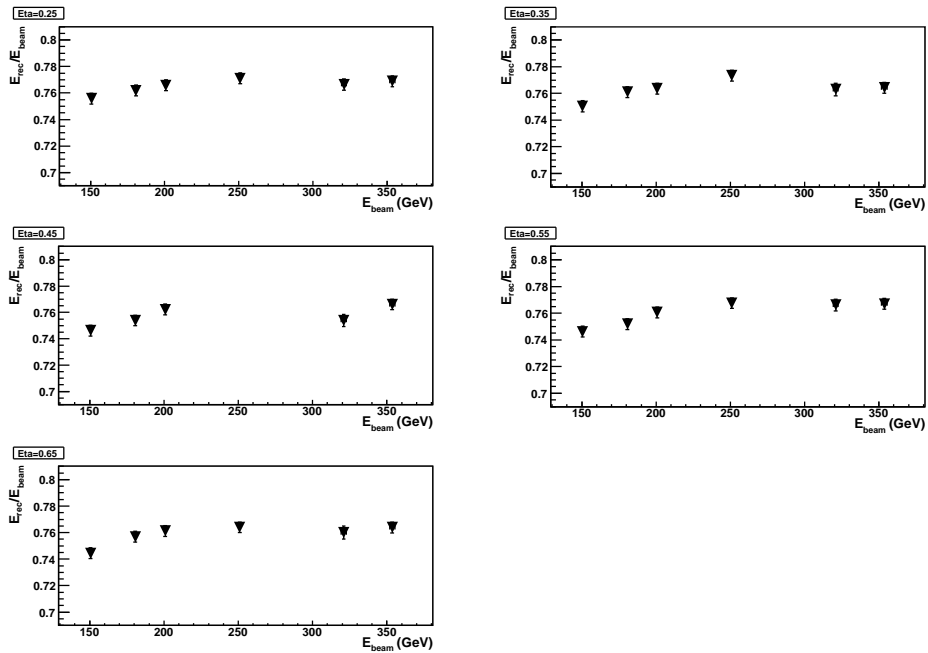}
  \caption{Combined response to pions for various $\eta$ positions.}
  \label{Response-combined}
\end{figure}

Figure~\ref{Response-combined} shows the ratio between the mean
reconstructed pion energy and the beam energy for different $\eta$
positions. Again the response is showing the typical behavior of a
non-comensating calorimeter : the response increases as the beam energy
increases (due to
the increase of the fraction of $\pi^0$ in the showers). The error on the
pion response is around 0.5\%. It comes from the uncertainty on the
beam energy. 

\section{Calorimeters response to very low energy pions\label{VLE}}
   \subsection{Motivations}

\begin{figure}[h]
  \includegraphics[height=.3\textheight]{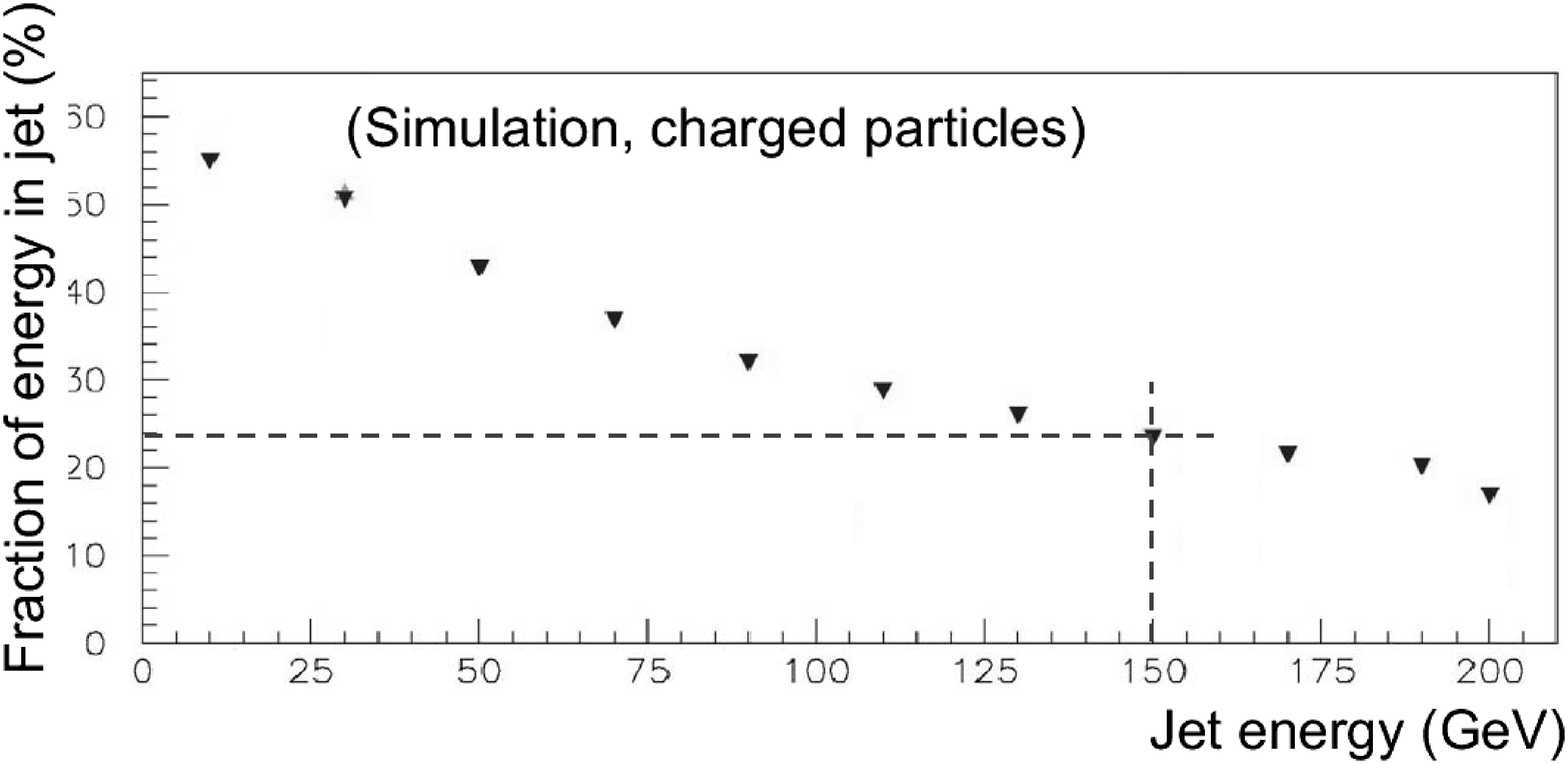}
  \caption{}
  \label{VLE-in-jets}
\end{figure}

In addition to the high energy program, very low energy data have been
taken during the 2004 combined testbeam. From the calorimetry point of view
this is motivated by the fact that very low energy particles (namely
E<10~GeV) have an important contribution in jets. Figure~\ref{VLE-in-jets}
shows a simulation of the fraction of energy brought by charged particles
with E<10~GeV as a function of the jet energy. One observes that about 25\% of
the energy in a 150~GeV jet is brought by particles with less than 10~GeV.
Therefore we need a good knowledge of the calorimeters response to very low
energy particles in order to check and tune the simulation of jets.

   \subsection{Selection of the very low energy pions}
Very low energy (VLE) data have been taken at $\eta=$0.2, 0.25, 0.35, 0.45, 0.55, 0.65
for energies between 1 and 9~GeV. The VLE beam is a mixture of low energy pions,
electrons and muons from pion decay. Due to the configuration of the beam line,
a large contamination by high energy muons is also expected.

\begin{figure}[h]
  \includegraphics[height=.2\textheight]{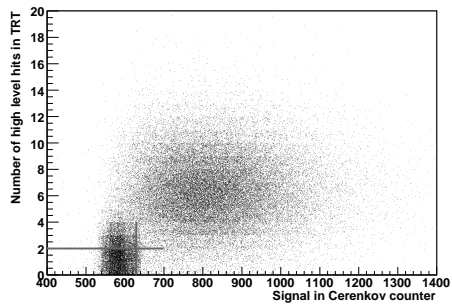}
  \caption{Correlation between the Cerenkov signal and the number of hits in
  TRT.}
  \label{VLE-epi-separation}
\end{figure}

The different particle separation has been performed using a Cerenkov
counter\footnote{Not an ATLAS detector.} and the Transition Radiation Tracker\footnote{TRT
is an ATLAS detector.} (TRT) that was in the beam line during the 2004 combined
test-beam. Both TRT and Cerenkov counter were used to separate pions from
electrons. The scatter plot in Figure~\ref{VLE-epi-separation} shows on the
x-axis the signal (in adc channel) in the Cerenkov counter and on the y-axis the
number of hits in TRT. In TRT, pions are expected to produced less transition
radiation X-ray than electrons. In the Cerenkov counter a lower signal is
expected for pions than for electrons. Electrons and pions (together with muons)
can be clearly identified on Figure~\ref{VLE-epi-separation}. The following cuts
have been applied to select pions : $Cerenkov<630~adc$ and {\it number of
hits$\le2$} (grey lines on Figure~\ref{VLE-epi-separation}).

\begin{figure}[h]
  \includegraphics[height=.2\textheight]{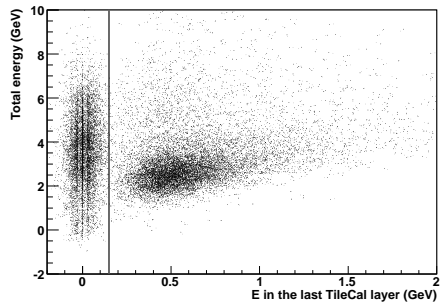}
  \caption{Correlation between the total energy and the energy in the last
  TileCal sample. The beam energy is 7~GeV.}
  \label{VLE-mupi-separation}
\end{figure}

Once electrons have been rejected we still have to deal with the contamination
of high energy muons. High energy
muons give a higher signal than very low energy pions in the last TileCal layer.
Indeed VLE pions are expected to lose most of their energy before reaching the end
of TileCal. High energy muons can be rejected asking less than a given energy 
in the last TileCal layer. Above 5~GeV this cut can introduce a bias because some
pions can reach this layer. Nevertheless such a cut is easy to reproduce in
the simulation. Figure~\ref{VLE-mupi-separation} shows the distribution of the total
energy as a function of the energy in the last TileCal layer. A large cloud with
a total energy around 2~GeV is observed. This is the expected signal for
high energy muons. The following cut has been chosen to select pions : {\it E
in the last TileCal layer < 150~MeV} (grey line in
Figure~\ref{VLE-mupi-separation}). Once pions have been separated from electrons
and muons, the energy $E_{raw}$ is determined fitting the total energy
distribution with a Gaussian function. $E_{raw}$ corresponds to the mean of the
function.
Again no corrections have been applied to take into
account the dead material effects.  

   \subsection{Calorimeters response to VLE pions}
\begin{figure}[h]
  \includegraphics[height=.3\textheight]{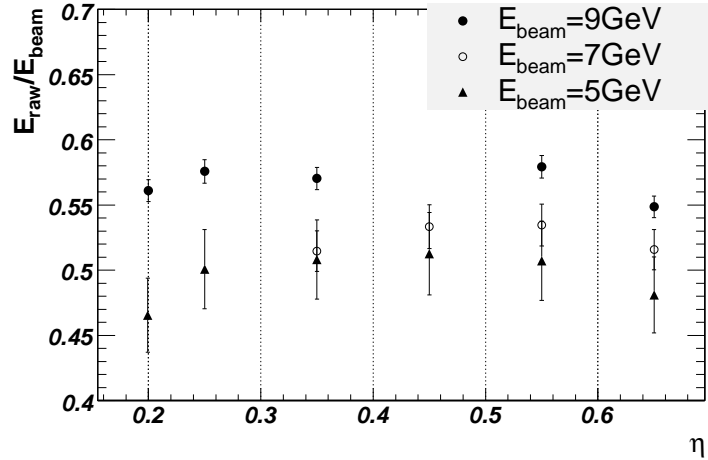}
  \caption{Response as a function of $\eta$ for very low energy pions.}
  \label{VLE-eta-scan}
\end{figure}

Very low energy pions have been selected as described above for various energies
and $\eta$ positions. Figure~\ref{VLE-eta-scan} gives the dependence of the
response ($E_{raw}/E_{beam}$) as a function of $\eta$. Due to the low number of
pions available, the uncertainty on the response is dominated by the statistical
error on the mean reconstructed energy. This statistical error varies from 1\%
at 9~GeV to 4\% at 5~GeV. The error on the knowledge of the beam energy is 
0.3\%. Systematic effects of the selection cuts have to be studied. 

\section{Conclusions and outlooks}
In this note results concerning the response of ATLAS central calorimeter
to pions have been presented. Data collected during a CERN testbeam were used.
The results were obtained studying :  
\begin{itemize}
   \item all high energy pions (energy larger than 50~GeV)
   \item high energy pions that start showering in TileCal
   \item very low energy pions (energy lower than 9~GeV)
\end{itemize}
The study of pions which start to produce a shower in TileCal has been performed
in order to determine the hadronic absolute scale for TileCal and to compare
TileCal performances in 2004 with the results obtained during 
previous standalone TileCal test-beams. From this analysis one can conclude that the resolution obtained in 2004 is
compatible with the values obtained during previous test-beams. The uncertainty
on TileCal response to pions is around 1\% (dominated by the statistical error). 

For the combined LAr+TileCal response to high energy pions, the uncertainty is
less than 1\%. It is dominated by the error on the beam energy.

For very low energy pions, the error on the LAr+TileCal response ranges from 1\%
at 9~GeV to 4\% at 5~GeV. Work is still in progress for the determination of the
systematic error introduced by the selection cuts (pion/electron separation,
pion/muon separation, beam profile...).

At the end of the talk possible jet calibration
strategy in ATLAS were discussed. The response of the
ATLAS calorimeters has been obtained (in the 2004 testbeam configuration). The
next step, which is already in progress, is to compare 
these results to the simulation in the same conditions (simulation of the
test beam geometry). If needed the simulation will be tuned according to the
testbeam results (amount of dead material, shower development...). 
Once a good agreement will be achieved, correction for dead
material, leakage, $e/\pi$ will be derived from simulation. Algorithms
could be proposed in order to improve the calorimeters performances for single
pions (like weighting techniques to improve linearity and resolution). 
The final objective is to obtain the jet absolute energy scale with a good
linearity and resolution.

%%%%%%%%%%%%%%%%%%%%%%%%%%%%%%%%%%%%%%%%%%%%%%%%
%% BACKMATTER
%%%%%%%%%%%%%%%%%%%%%%%%%%%%%%%%%%%%%%%%%%%%%%%%



\bibliographystyle{aipproc}   % if natbib is available

%%\bibliography{sample}

%  \typeout{******************************************}
%  \typeout{** Please run "bibtex \jobname" to optain}
%  \typeout{** the bibliography and then re-run LaTeX}
%  \typeout{** twice to fix the references!}
%  \typeout{******************************************}

%\end{document}

%%%%%%%%%%%%%%%%%%%%%%%%%%%%%%%%%%%%%%%%%%%
%% The following lines show an example how to produce a bibliography
%% without the help of the BibTeX program. This could be used instead
%% of the above.
%%%%%%%%%%%%%%%%%%%%%%%%%%%%%%%%%%%%%%%%%%%

%\endinput
%%
\end{document}